\documentclass[useAMS,usenatbib]{mn2e}
\usepackage{graphicx}
\title[Escapees from low-$N$ clusters]{Escaping stars from young
  low-$N$ clusters}
\author[C.~Weidner, I.~A.~Bonnell and N.~Moeckel]
{C.~Weidner$^{1}$\thanks{E-mail: 
    cw60@st-andrews.ac.uk}, I.~A.~Bonnell$^{1}$\thanks{E-mail:  
    iab1@st-andrews.ac.uk} and N.~Moeckel$^{1,2}$\thanks{E-mail:
    moeckel@ast.cam.ac.uk}\\
$^{1}$Scottish Universities Physics Alliance (SUPA), School of Physics and
  Astronomy, University of St. Andrews, North Haugh,\\
 St. Andrews, Fife KY16 9SS, UK\\
$^{2}$Institute of Astronomy, Madingley Road, Cambridge CB3 0HA, UK
}

\begin{document}
\bibliographystyle{mnras}
\date{Accepted . Received 2010; in original form }

\pagerange{\pageref{firstpage}--\pageref{lastpage}} \pubyear{2010}

\maketitle

\label{firstpage}

\begin{abstract}
With the use of $N$-body calculations the amount and properties of
escaping stars from low-$N$ ($N$ = 100 and 1000) young embedded star
clusters prior to gas expulsion are studied over the first 5 Myr
of their existence. Besides the number of stars also different initial
radii and binary populations are examined as well as virialised and
collapsing clusters. It is found that these clusters can loose
substantial amounts (up to 20\%) of stars within 5 Myr with
considerable velocities up to more than 100 km/s. Even with their mean
velocities between 2 and 8 km/s these stars will still be travelling
between 2 and 30 pc during the 5 Myr. Therefore can large amounts of
distributed stars in star-forming regions not necessarily be counted
as evidence for the isolated formation of stars.
\end{abstract}


\section{Introduction}
\label{se:intro}
Stars do not form in absolute isolation but rather in groups or
embedded clusters in dense molecular cloud cores
\citep{LL95,LL03,AM01,AMG07}. Theoretical and observational results
indicate that these groups and the majority of the
embedded clusters dissolve quickly due to gas expulsion after about 10
Myr and release their stars into the galactic field
\citep{KAH01,KB03,AM01,LL03,WKNS06,BK08}. But the ratio of the
distributed mode of star formation and the clustered one is not fully
understood. Observations show a distributed fraction of stars in the
Orion molecular clouds of about 20\% \citep{AMG07}. It is generally
assumed that these fraction is primordial as dynamical evolution of
the clusters is considered to slow to reproduce the observed
fraction of isolated stars at such a young stage. But for a full
physical understanding of the formation of stars it is vital to know
all modes of star formation in as much detail as possible.
The purpose of this work is therefore to access whether or not the
observed amount of distributed stars outside embedded clusters
could have a dynamical origin and henceforth being initially formed
in the young, tight embedded clusters or if there is intrinsically
isolated star formation. Especially in star-forming regions which do
not show massive star clusters the amount of stars released through the
dynamical evolution of low-$N$ clusters is not well known.
In order to study this question a large series of $N$-body
calculations is performed and the number, the velocity and the mass
spectrum of stars ejected due to dynamical interactions within 5 Myrs
from this clusters are studied.

\section{The cluster setup}
\label{sec:setup}
With the use of the $N$-body6 code \citep{SJA99b,SJA99,SJA03,SJA08}
the evolution of numerical star clusters with small numbers of stars
are examined. Two different numbers of stars, $N$, are used, 100 and 
1000. As the actual number of escaped stars can be rather low each set
of initial conditions is simulated 100 times using different random
number seeds to set up the masses, velocities and positions of the
stars in order to improve the statistical significance of the
results. All stars in the clusters follow the canonical IMF \citep{Kr02}.

Besides the initial number of stars several other initial parameters
are studied with each 100 calculations. 
\begin{itemize}
  \item All stars are initially single stars,
  \item all stars reside in binaries,
  \item 50\% of the stars are in binaries,
  \item the cluster is initially in virial equilibrium ,
  \item the cluster is initially sub-virial (collapsing with the
    kinetical energy being one-tenth of the potential energy).
  \item All star clusters are setup as Plummer-spheres
    \citep{Pl11} with different initial half-mass radii of 0.1,
    0.25 or 0.5 pc.
\end{itemize}

Each of these setups are then numerically evolved with $N$-body6 for 5
Myr\footnote{Depending on the initial radius of the cluster 5 Myr are
  between about 10 and 150 crossing times. See eq.~\ref{eq:tcross} for
how to calculate the crossing time.}. The evolution time of 5
Myr is chosen to minimise mass loss due to stellar evolution, to avoid
supernovae and primarily to have a setting of a
star cluster still embedded in its parental cloud, prior to gas
expulsion. Though, clusters initially probably form from clumpy
structures \citep{WBM00,CH08}. Simulations of sub-virial clumpy initial
conditions \citep{AGP09} have shown that such clusters mass segregate
faster than virial clusters due to the formation of short-lived very
dense cores. But these calculations did not include a background gas
potential.

To emulate the gas in which the stars are embedded during
this time, the cluster is setup with an additional plummer potential
\citep{Pl11,BT87} of the same half-mass radius and the same mass
as the cluster. Thus assuming a star-formation efficiency (SFE) of
50\%. Though, global SFE's are observationally around 30\%
\citep{LL03}, locally within the region where the cluster actually
forms the value is most likely higher \citep{MB10}. A SFE of 50\% has
therefore been chosen. A large differences in results between 30 and
50\% SFE is not to be expected. The gas potential is in all cases kept
constant during the whole calculation time.
As the results are qualitatively similar to $N$-body studies
without a gas background potential \citep[e.g. ][]{K98,KB03}, the exact
shape of the potential does not seems to be overly important. But
detailed further studies are needed to clarify the full impact of the
shape and depth of the potential on the properties of escaping stars.

The decision to include primordial binaries into the study is based on
the observational results that open clusters host large number of
binaries \citep{SGE09,SCB09}. The initial setup of the binary
properties increase the total parameter space of the calculations
significantly. As a cluster of the richness of the Orion Nebula
cluster stellar dynamics may change the mass function and binary
properties in the core significantly in less than $10^{6}$ yr
\citep{GB06,PAK06,PO07,AGP09}. And as the majority of the present-day field
binaries are being processed through star clusters, their properties
are not suited to be used for the numerical setup
\citep{DM91,Kr95a,Kr95b,Kr95c,Kr95d,KB03}. Therefore, the results of
the inverse dynamical population synthesis of \citet{Kr95c} are used,
which predict a flat (thermal) period distribution (resulting in a
uniform logarithmic semi-major axis distribution) and random pairing
of the stars in systems of low-mass ($<$ 1 $M_\odot$) stars
\citep{Kr95c,Kr95d}. While there are indications for non-random
pairing of massive stars \citep{KF07,WK07c}, the low-mass results are
also used for massive binaries in this study. Firstly, because the
mass where a change of properties might occur is not determined yet
and it is not clear if mass-ratios near to unity for massive stars are
primordial or already a sign of dynamical evolution \citep{PO07} and
secondly, because due to the choice of $N$ = 100 and 1000, only a
relatively small number of massive stars is to be expected anyway. For
example, there should be only three stars above 8 $M_\odot$ for $N$ =
1000 and a Kroupa-IMF.

It should be noted here, that rapid changes in the binary properties
might also at least partly due to the initial conditions of 
$N$-body calculations and do not really reflect the star formation
process. I.e.~that unstable binaries that are easily destroyed in
early cluster dynamics may not even form in the first place
\citep{MB10}.

In the $N$ = 1000 cases, for the IMF a maximal mass, $m_\mathrm{max}$,
of 25 $M_\odot$ is chosen, in accordance with the
$m_\mathrm{max}-M_\mathrm{ecl}$-relation by \citet{WKB09}. In the $N$
= 100 case this relation yields a $m_\mathrm{max}$ of 7 $M_\odot$. But
also comparison calculations are done using $m_\mathrm{max}$ = 25
$M_\odot$ for the $N$ = 100 case.

\section{Results}
\label{se:disc}
\begin{figure}
\begin{center}
\includegraphics[width=8cm]{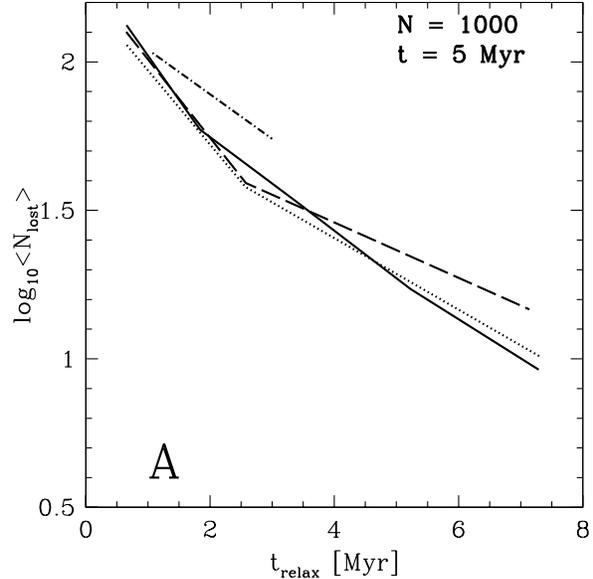}
\includegraphics[width=8cm]{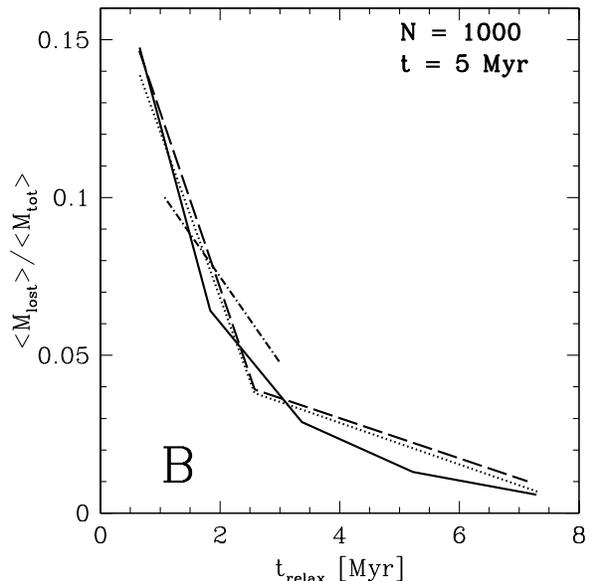}
\vspace*{-2.0cm}
\caption{Mean number of stars lost vs $t_\mathrm{relax}$ ({\it Panel
    A}) and mean percentage of cluster mass lost vs $t_\mathrm{relax}$
  ({\it Panel B}). The {\it solid line} shows the case
  without any initial binaries, the {\it dotted line} is the 50\%
  initial binary case, the {\it dashed line} is the 100\% initial
  binaries case and the {\it dashed-dotted line} marks the sub-virial case.}
\label{fig:N1000restr1}
\end{center}
\end{figure}

\begin{figure}
\begin{center}
\includegraphics[width=8cm]{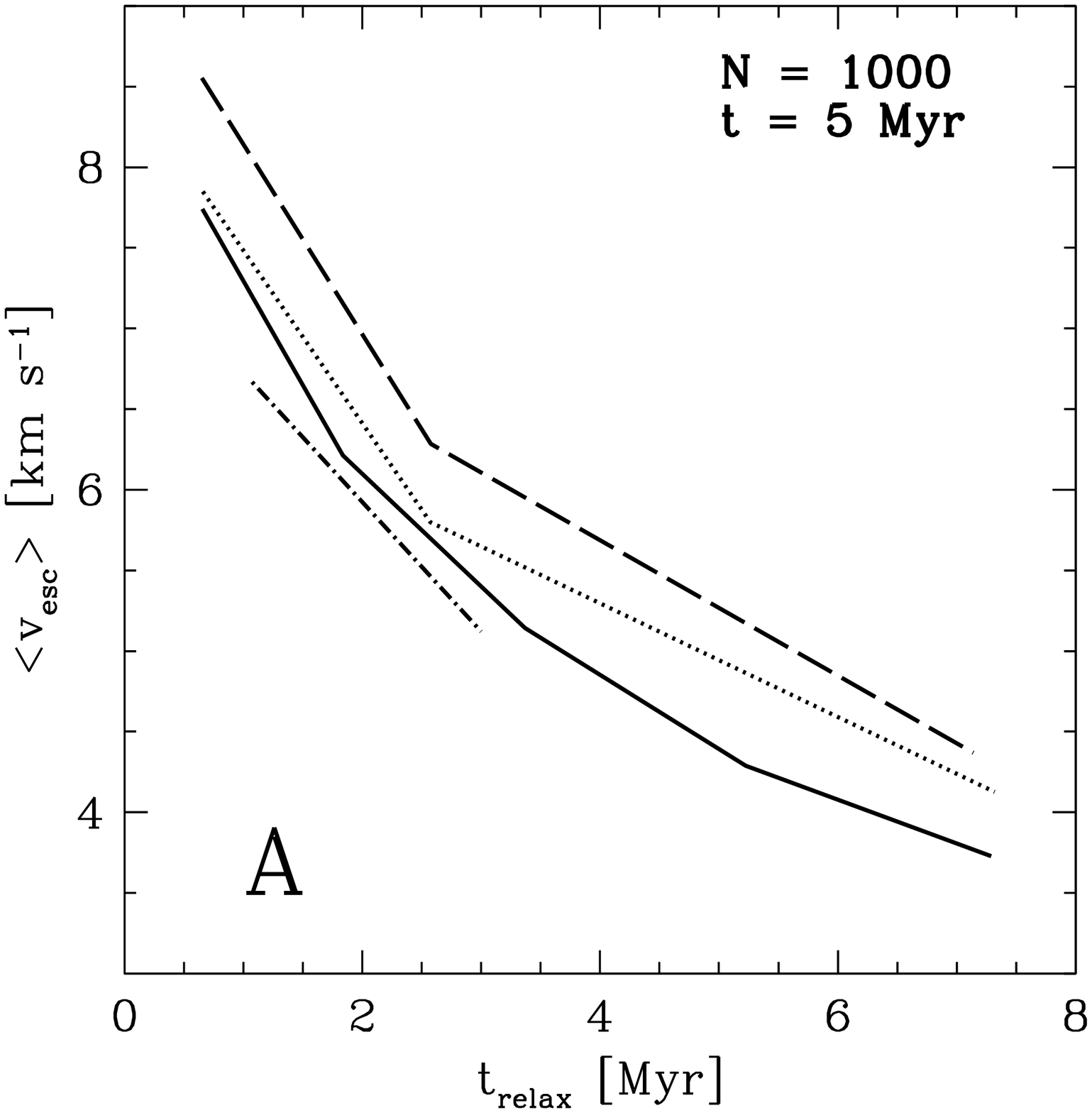}
\includegraphics[width=8cm]{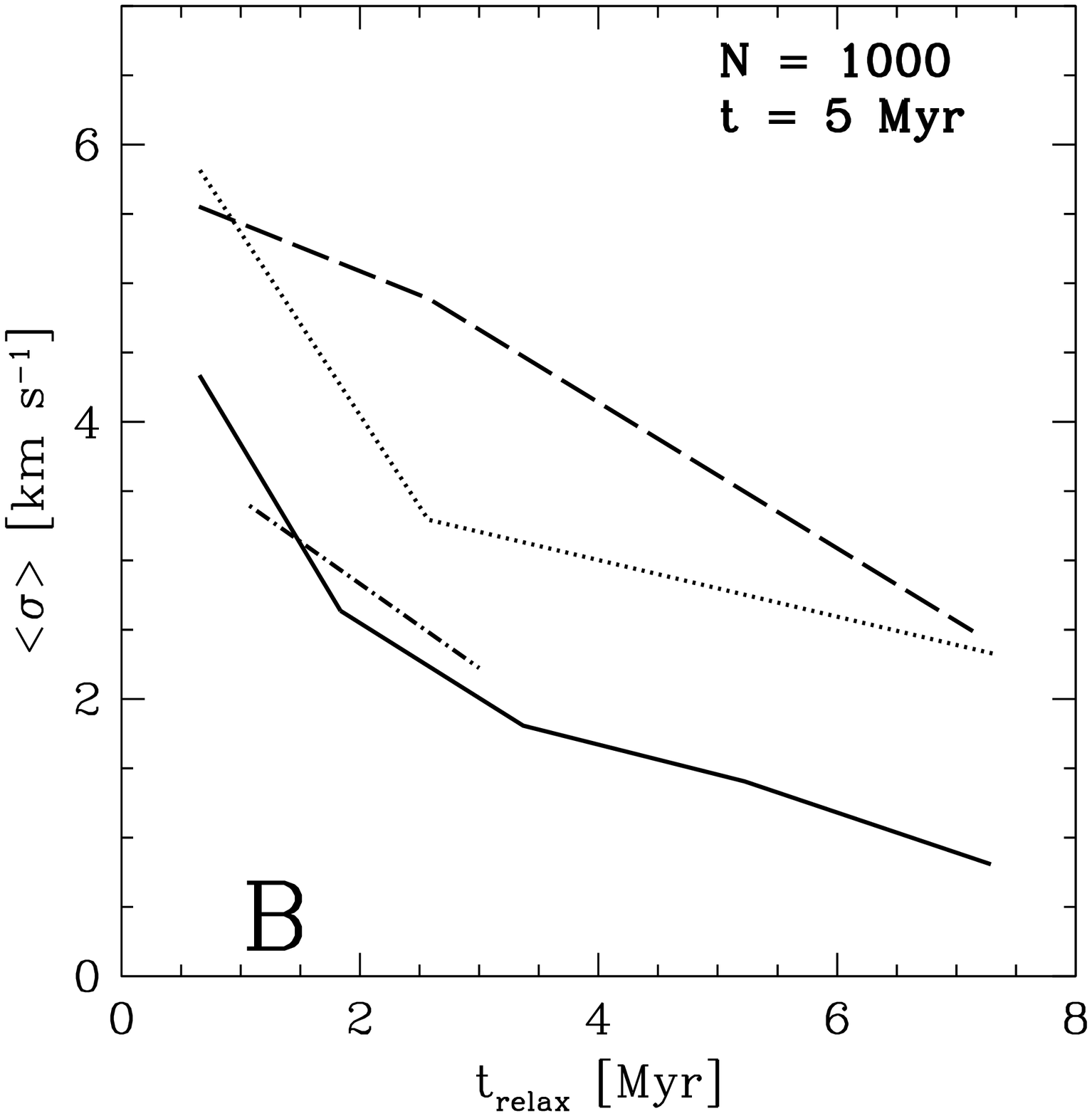}
\vspace*{-2.0cm}
\caption{Mean escape velocity, $<$$v_\mathrm{esc}$$>$, vs
  $t_\mathrm{relax}$ ({\it Panel A}) and mean velocity dispersion,
  $<$$\sigma$$>$, vs $t_\mathrm{relax}$ ({\it Panel B}). The {\it
    solid line} shows the case without any initial binaries, the {\it
    dotted line} is the 50\% initial binary case, the {\it dashed
    line} is the 100\% initial binaries case and the {\it
    dashed-dotted line} marks the sub-virial case.}
\label{fig:N1000restr2}
\end{center}
\end{figure}

\begin{figure}
\begin{center}
\includegraphics[width=8cm]{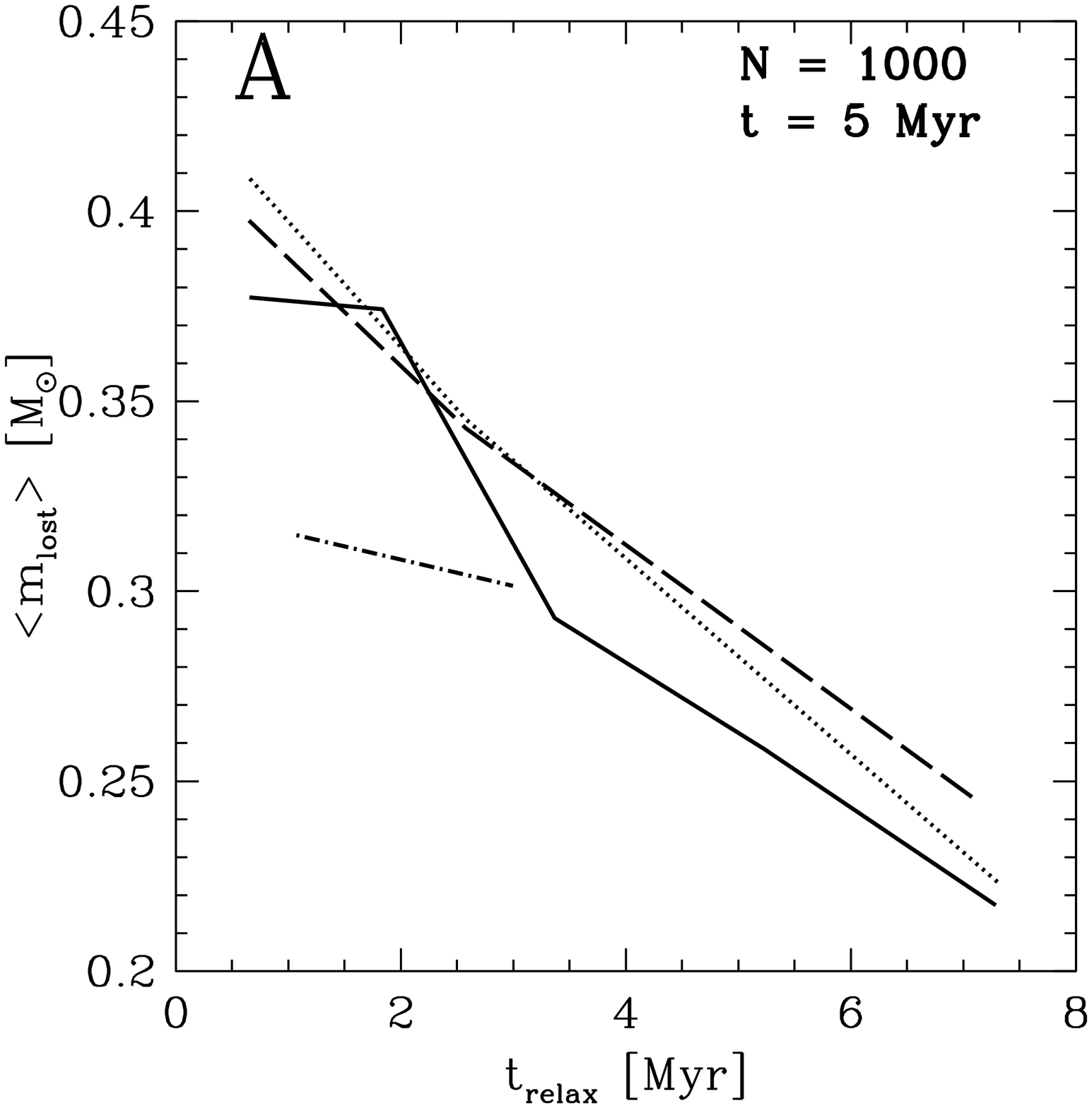}
\includegraphics[width=8cm]{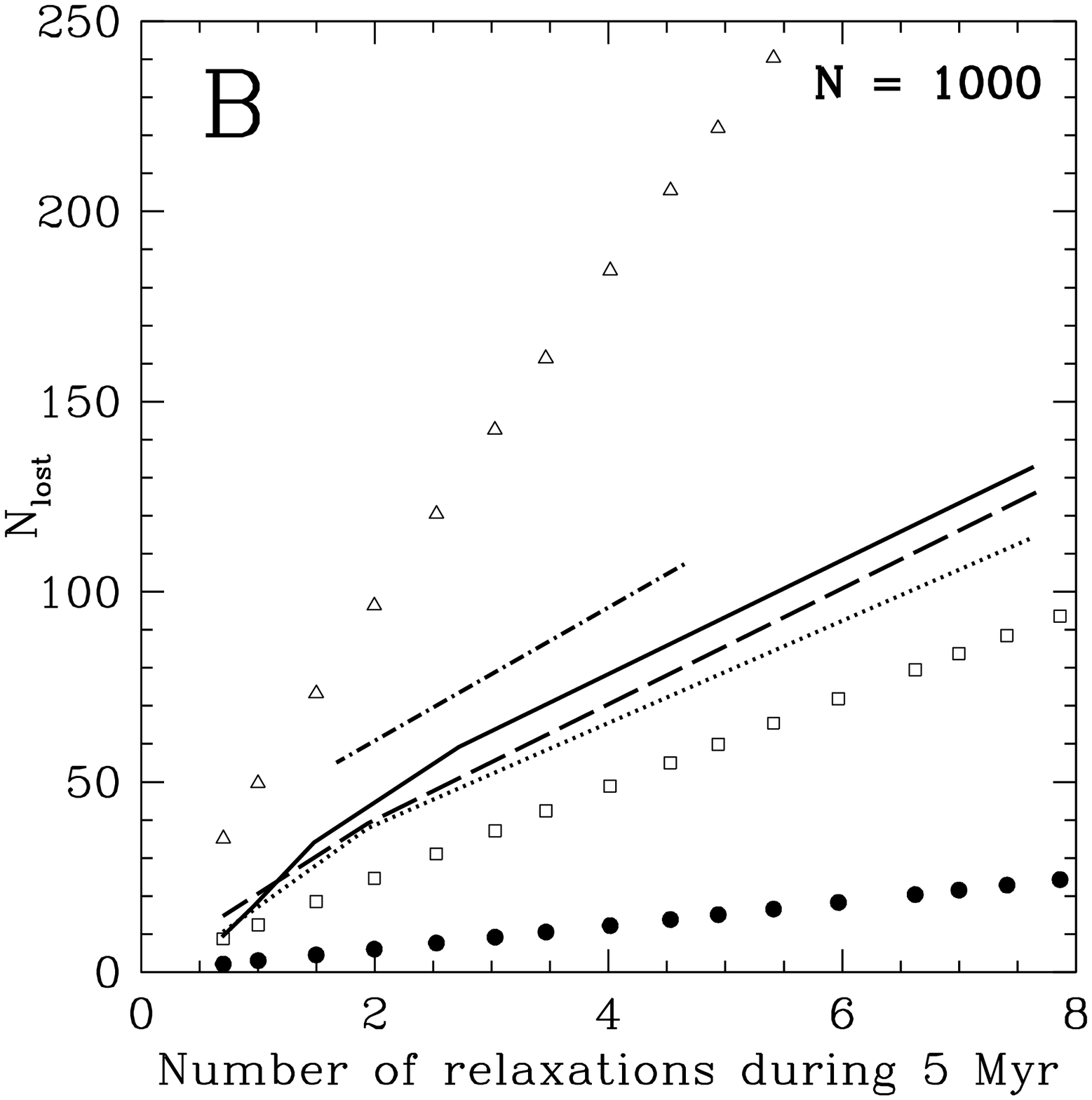}
\vspace*{-2.0cm}
\caption{{\it Panel A:} Mean mass of the escaped stars vs
  $t_\mathrm{relax}$. The {\it solid line} shows the case without any 
  initial binaries, the {\it dotted line} is the 50\% initial binary
  case, the {\it dashed line} is the 100\% initial binaries case and
  the {\it dashed-dotted line} marks the sub-virial case. {\it Panel
    B:} Total number of lost stars vs number of relaxation times the
  cluster experienced during the 5 Myr simulation time. The {\it solid
    line} shows the case without any initial binaries, the {\it dotted
    line} is the 50\% initial binary case, the {\it dashed line} is
  the 100\% initial binaries case and the {\it dashed-dotted line}
  marks the sub-virial case. The {\it solid black dots} are the
  predicted numbers from eq.~\ref{eq:Nloss}, the {\it open squares}
  from eq.~\ref{eq:Ntdiss} and the {\it open triangles} from
  eq.~\ref{eq:Ntevap}.}
\label{fig:sres2b}
\end{center}
\end{figure}

\begin{figure}
\begin{center}
\includegraphics[width=8cm]{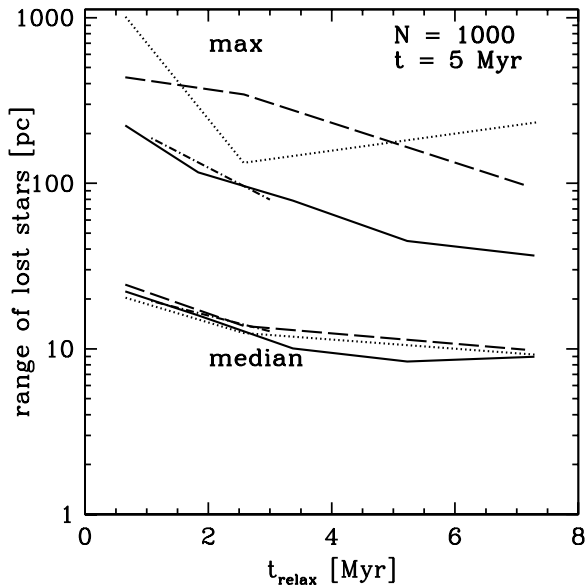}
\vspace*{-2.0cm}
\caption{Median and maximal flight ranges from the cluster centre of
  the escaped stars vs $t_\mathrm{relax}$. The {\it solid line} shows
  the case without any initial binaries, the {\it dotted line} is the
  50\% initial binary case, the {\it dashed line} is the 100\% initial
  binaries case and the {\it dashed-dotted line} marks the sub-virial case.}
\label{fig:N1000range}
\end{center}
\end{figure}

\begin{figure}
\begin{center}
\includegraphics[width=8cm]{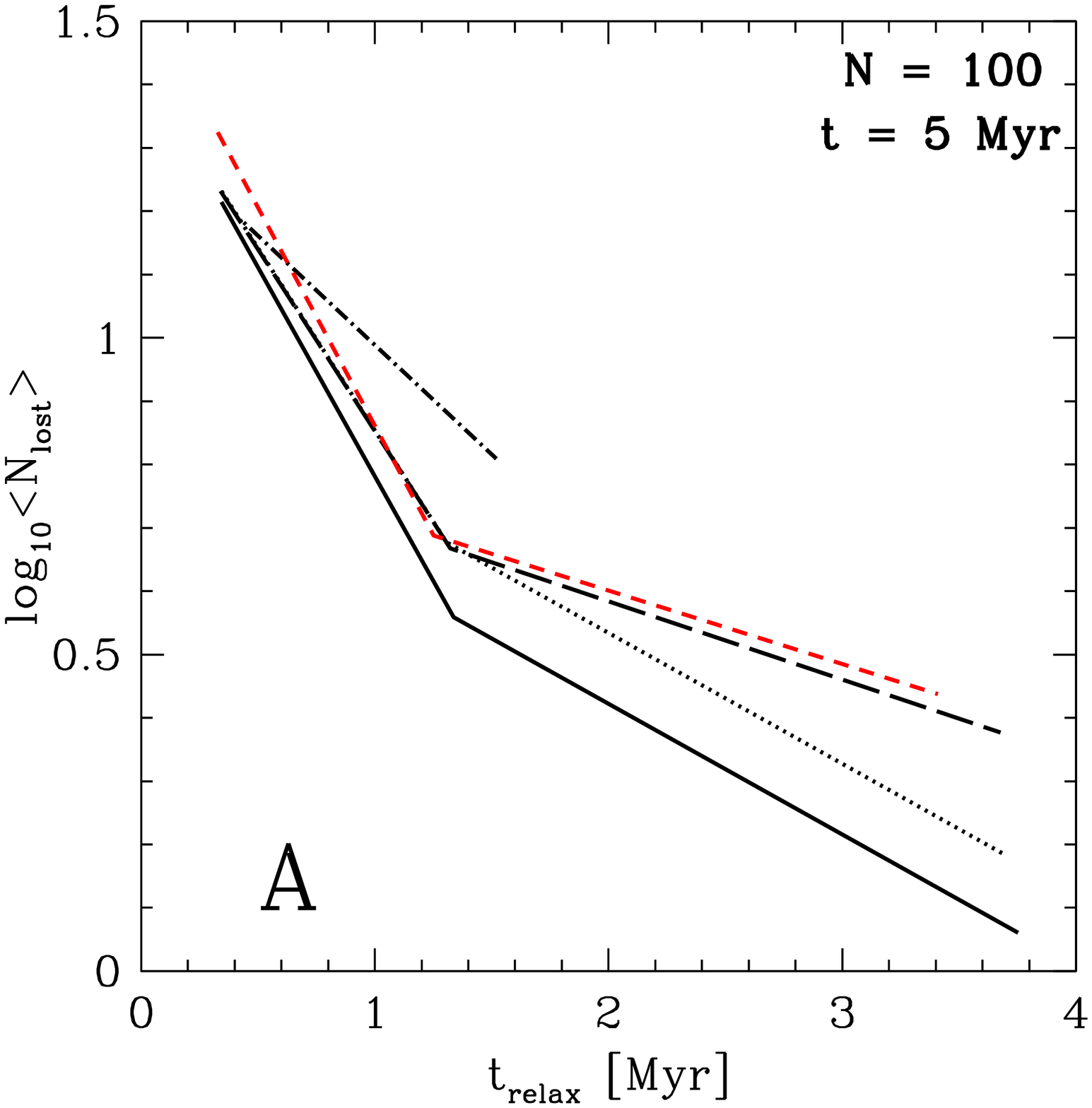}
\includegraphics[width=8cm]{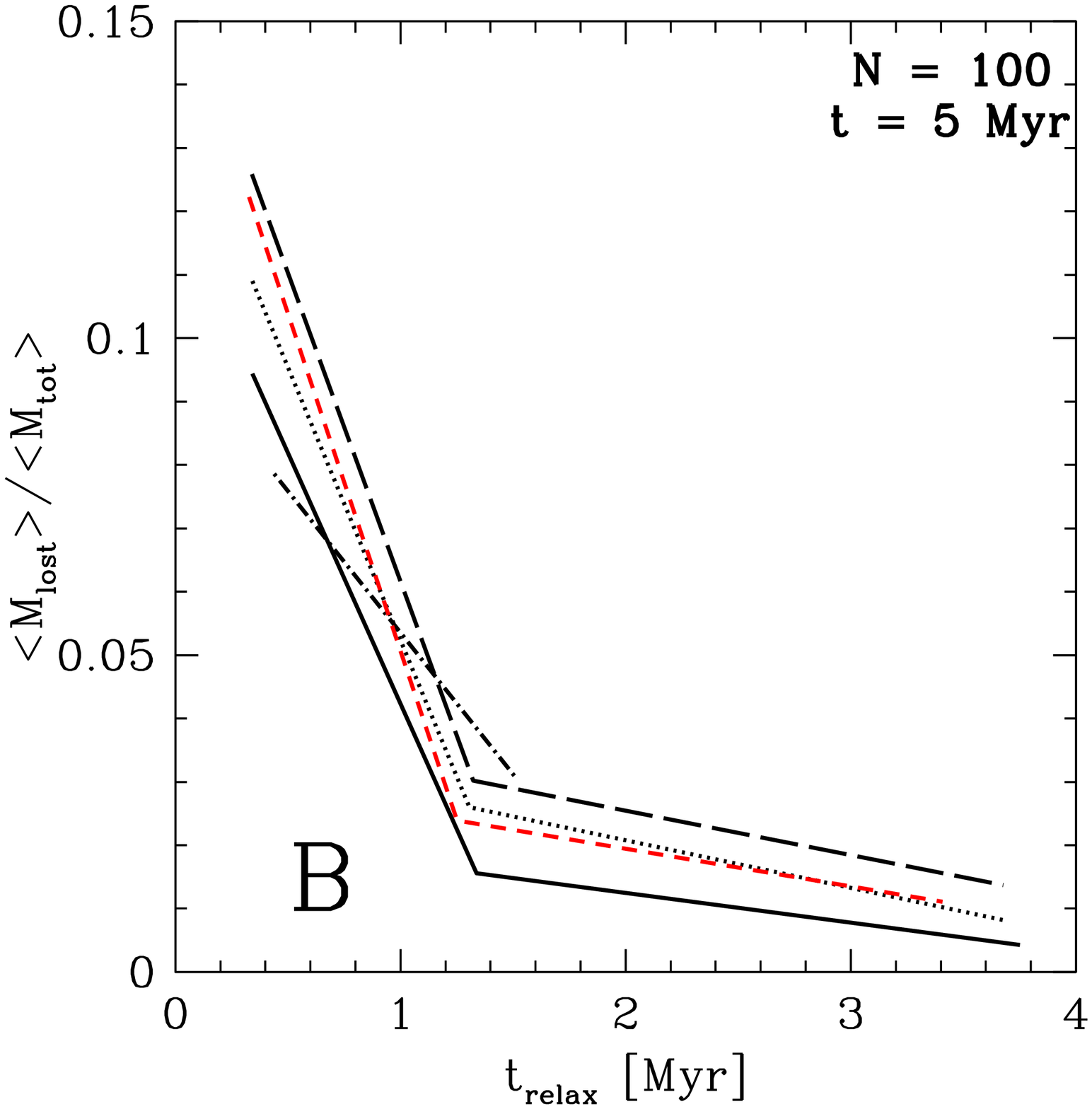}
\vspace*{-2.0cm}
\caption{Mean number of stars lost vs $t_\mathrm{relax}$ ({\it Panel
    A}) and mean percentage of cluster mass lost vs $t_\mathrm{relax}$
  ({\it Panel B}). The {\it solid line} shows the case
  without any initial binaries, the {\it dotted line} is the 50\%
  initial binary case, the {\it long-dashed line} is the 100\% initial
  binaries case, the {\it dashed-dotted line} marks the sub-virial case
  and {\it short-dashed line} the case with $m_\mathrm{max}$ = 25 $M_\odot$.}
\label{fig:N100restr1}
\end{center}
\end{figure}

\begin{figure}
\begin{center}
\includegraphics[width=8cm]{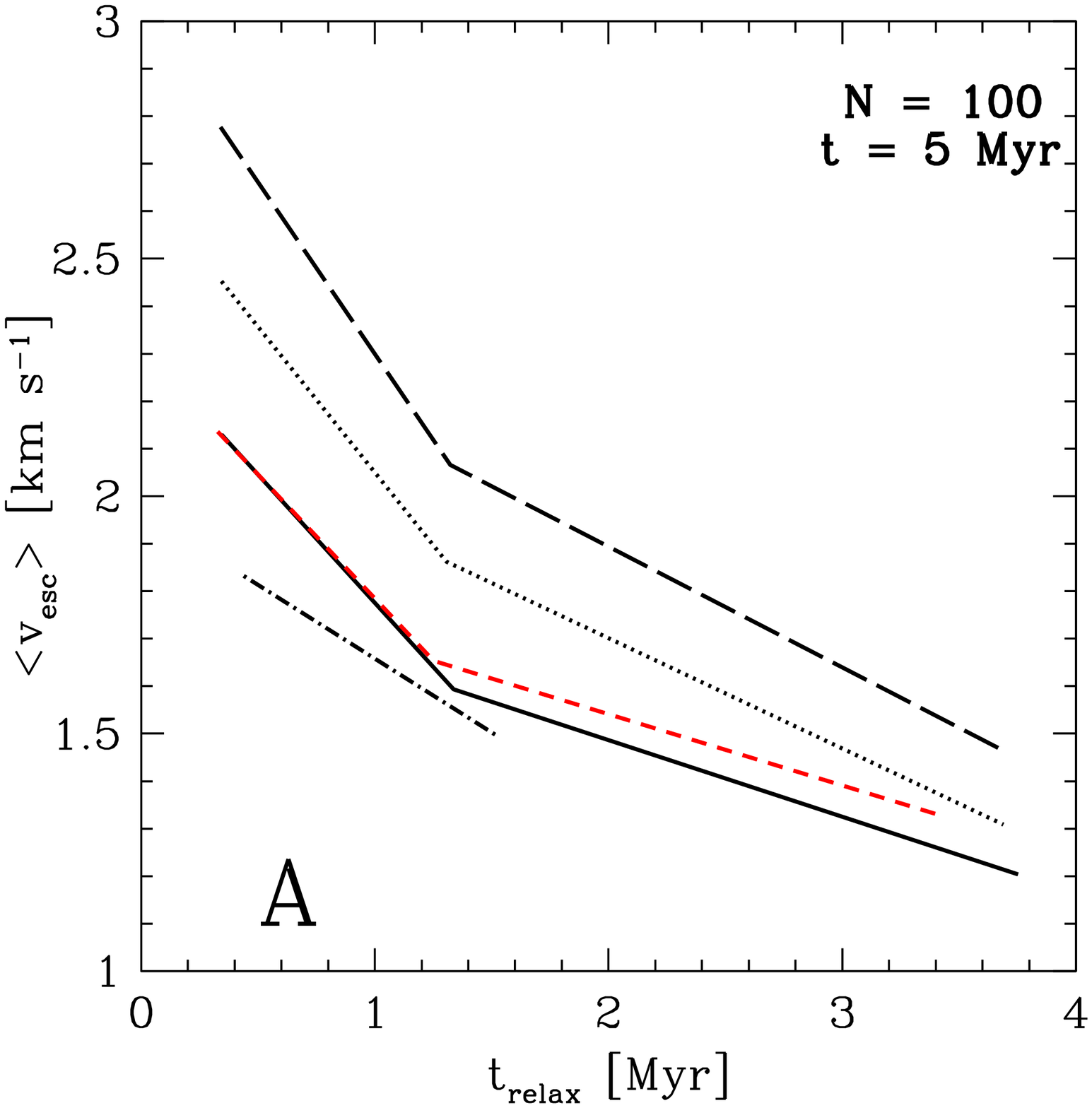}
\includegraphics[width=8cm]{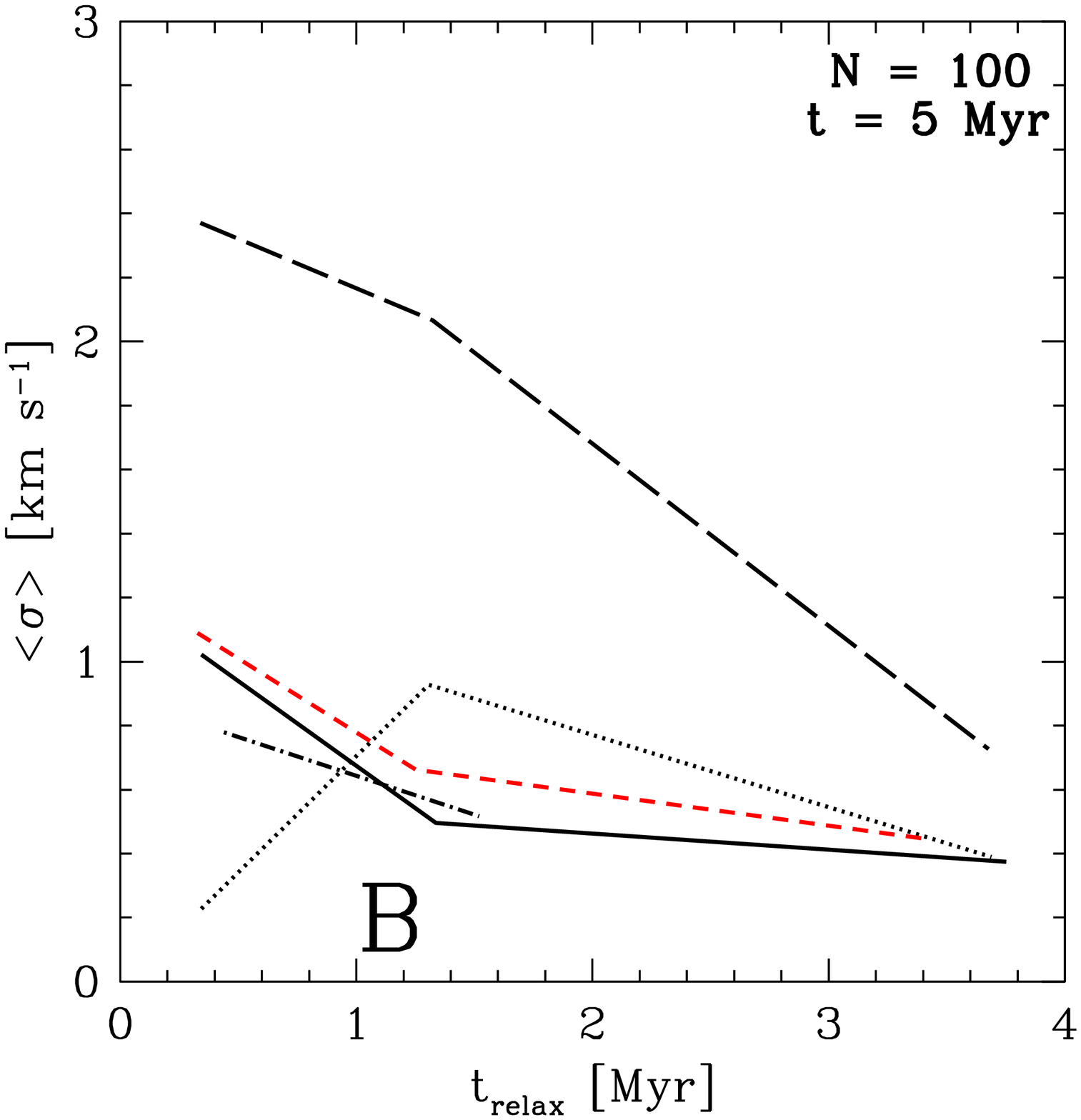}
\vspace*{-2.0cm}
\caption{mean escape velocity vs $t_\mathrm{relax}$ ({\it Panel A})
  and mean velocity dispersion vs $t_\mathrm{relax}$ ({\it Panel
    B}). The {\it solid line} shows the case without any initial
  binaries, the {\it dotted line} is the 50\% initial binary case, the
  {\it long-dashed line} is the 100\% initial binaries case, the {\it
    dashed-dotted line} marks the sub-virial case and {\it
    short-dashed line} the case with $m_\mathrm{max}$ = 25 $M_\odot$.}
\label{fig:N100restr2}
\end{center}
\end{figure}

\begin{figure}
\begin{center}
\includegraphics[width=8cm]{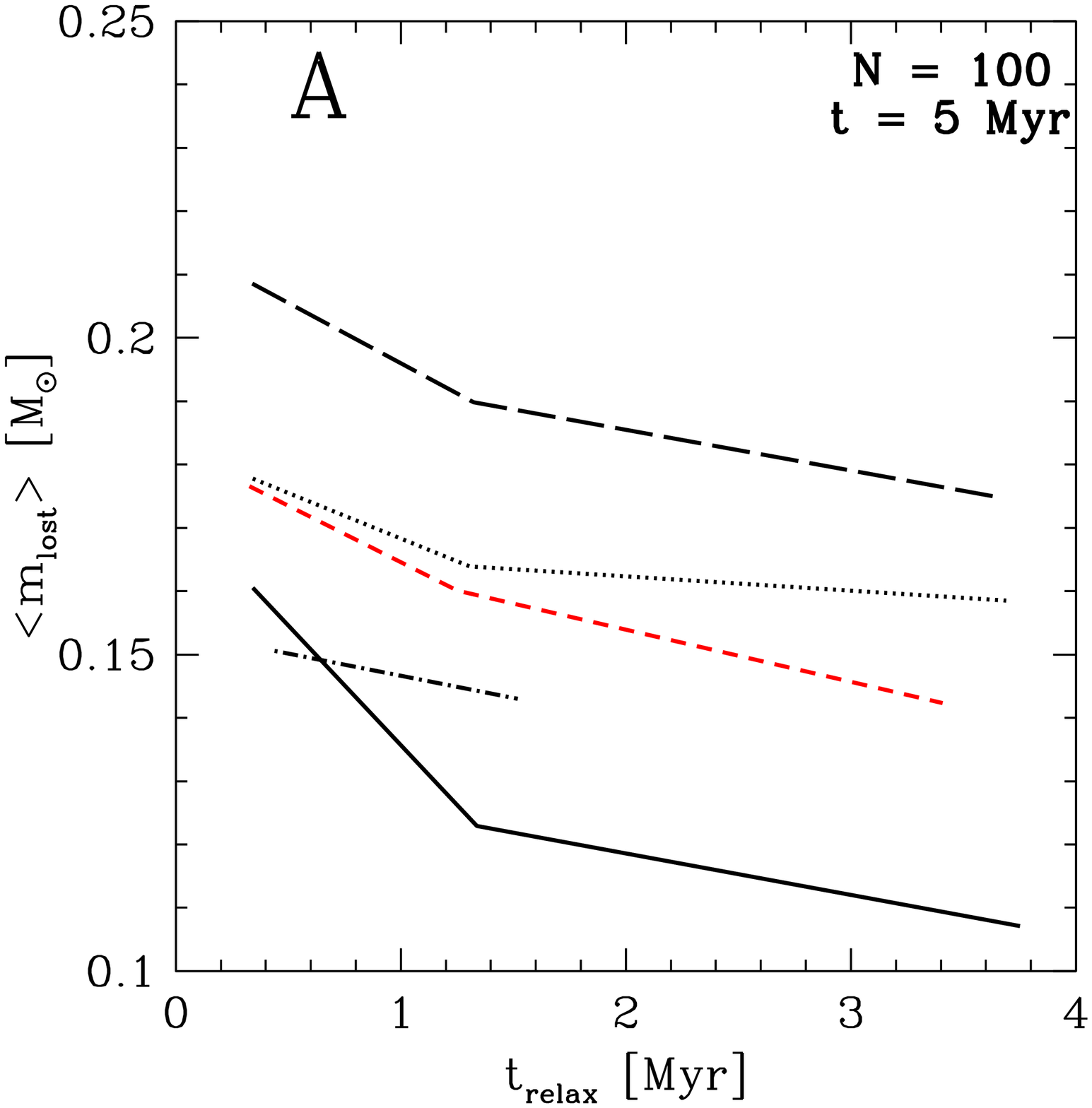}
\includegraphics[width=8cm]{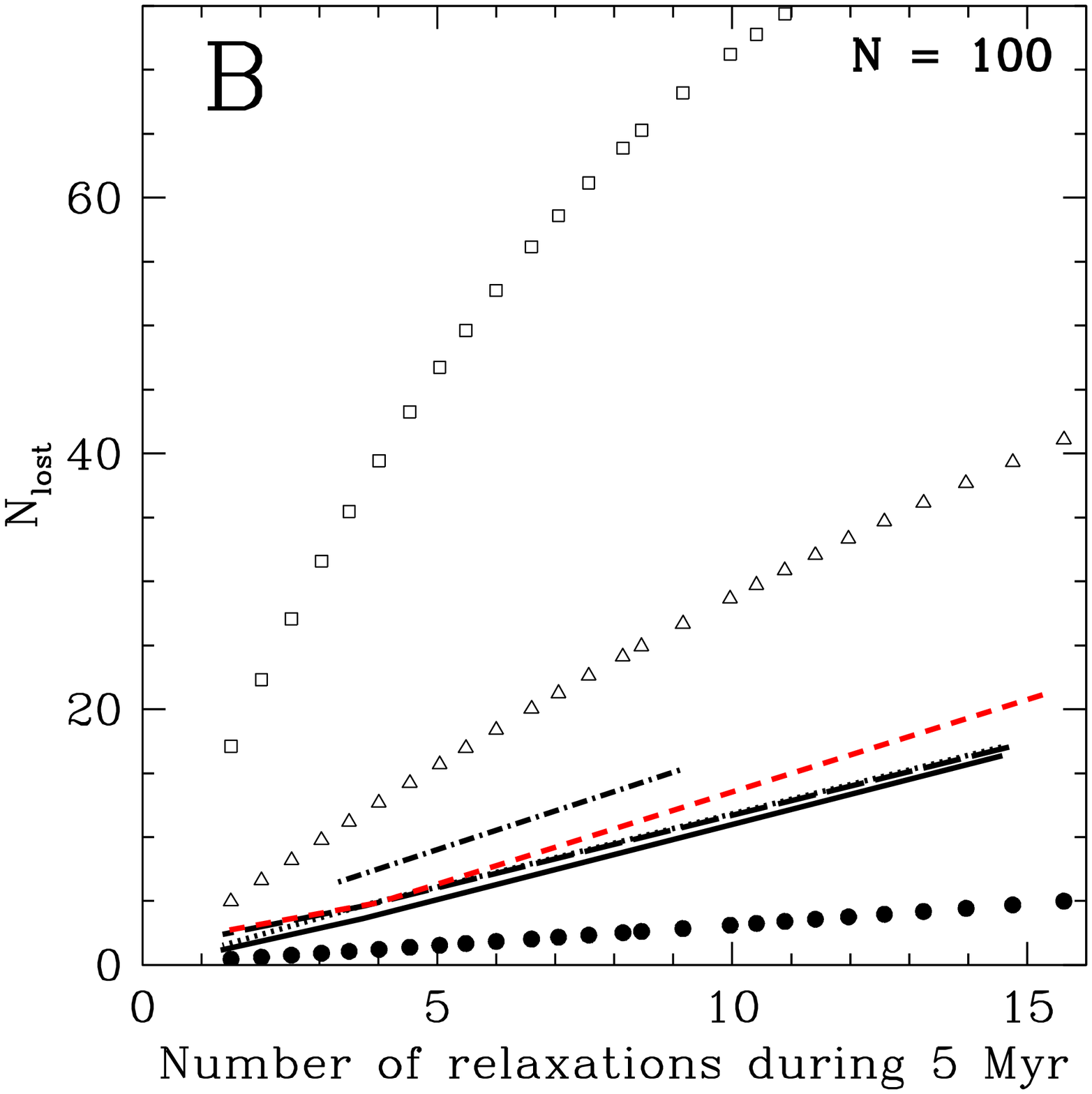}
\vspace*{-2.0cm}
\caption{{\it Panel A:} Mean mass of the escaped stars vs
  $t_\mathrm{relax}$. The {\it solid line} shows the case without any
  initial binaries, the {\it dotted line} is the 50\% initial binary
  case, the {\it long-dashed line} is the 100\% initial binaries case,
  the {\it dashed-dotted line} marks the sub-virial case and {\it
    short-dashed line} the case with $m_\mathrm{max}$ = 25
  $M_\odot$. {\it Panel B:} Total number of lost stars vs number of
  relaxation times the 
  cluster experienced during the 5 Myr simulation time. The {\it solid
    line} shows the case without any initial binaries, the {\it dotted
    line} is the 50\% initial binary case, the {\it long-dashed line} is
  the 100\% initial binaries case, the {\it dashed-dotted line}
  marks the sub-virial case and {\it short-dashed line} the case with
  $m_\mathrm{max}$ = 25 $M_\odot$. The {\it solid black dots} are the
  predicted numbers from eq.~\ref{eq:Nloss}, the {\it open squares}
  from eq.~\ref{eq:Ntdiss} and the {\it open triangles} from
  eq.~\ref{eq:Ntevap}.}
\label{fig:sres5b}
\end{center}
\end{figure}

\begin{figure}
\begin{center}
\includegraphics[width=8cm]{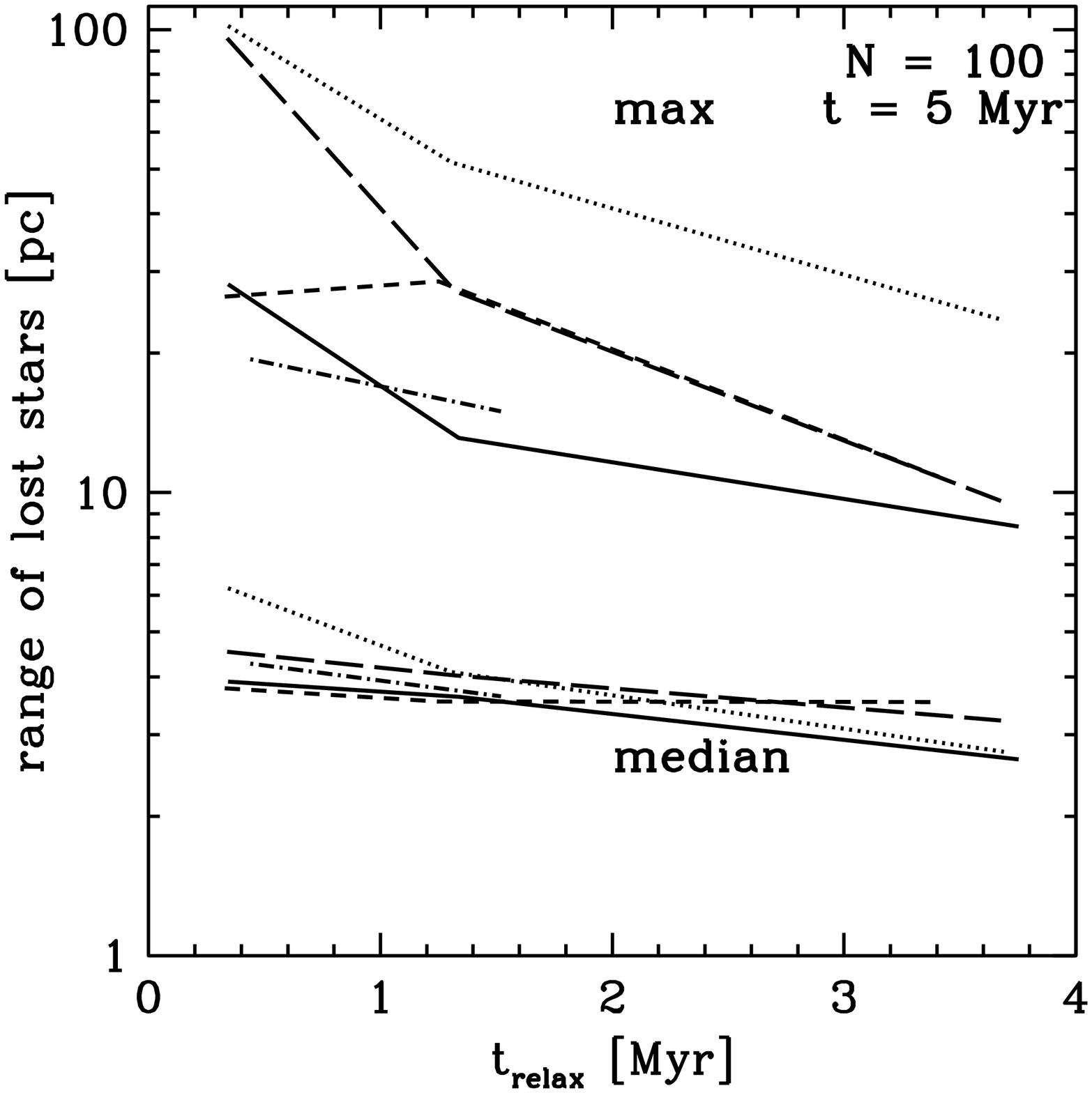}
\vspace*{-2.0cm}
\caption{Median and maximal flight ranges from the cluster centre of
  the escaped stars vs $t_\mathrm{relax}$. The {\it solid line} shows
  the case without any initial binaries, the {\it dotted line} is the
  50\% initial binary case, the {\it dashed line} is the 100\% initial
  binaries case, the {\it dashed-dotted line} marks the sub-virial
  case and {\it short-dashed line} the case with
  $m_\mathrm{max}$ = 25 $M_\odot$.}
\label{fig:N100range}
\end{center}
\end{figure}

In the series of Figs.~\ref{fig:N1000restr1} to \ref{fig:N100range} the
results of the $N$-body6 calculations after 5 Myr are presented. There the
mean number of lost stars, $<$$N_\mathrm{lost}$$>$, the mean
lost fraction of mass, $<$$M_\mathrm{lost}$$>$/$<$$M_\mathrm{tot}$$>$, the mean
escape velocity, $<$$v_\mathrm{esc}$$>$, the mean escape velocity
dispersion, $<$$\sigma$$>$, and the mean mass of the lost stars,
$<$$m_\mathrm{lost}$$>$, are studied in dependence of the initial cluster
relaxation time, $t_\mathrm{relax}$. The latter is calculated by
using the following equations from \citet{BT87}:
\begin{equation}
\label{eq:trelax}
t_\mathrm{relax} \approx \frac{N}{8 \ln{N}} t_\mathrm{cross},
\end{equation}
with $N$ being the number of stars and $t_\mathrm{cross}$ the crossing
time a star needs to travel through the cluster which is given by,
\begin{equation}
\label{eq:tcross}
t_\mathrm{cross} = 2 \sqrt{\frac{R_\mathrm{ecl}^3}{GM_\mathrm{tot}}},
\end{equation}
with $G$ being Newton's gravitational constant, $R_\mathrm{ecl}$ the
cluster radius and $M_\mathrm{tot}$ its mass. 

It is important to note that the relaxation time of a cluster is
not a constant but also evolves with time. Though, for the clusters
which start in virial equilibrium, the change is negligible over the
short period of time studied here. The clusters expand about
$\sim$10\% due to evolutionary and dynamical mass loss. This leads to
a $\sim$10\% increase in $t_\mathrm{relax}$. But as the sub-virial
clusters contract significantly during the calculation time span,
their $t_\mathrm{relax}$ decreases and becomes comparable to a cluster
with half their half-mass radius.

In the {\it B panels} of the Figs.~\ref{fig:sres2b} and
\ref{fig:sres5b} are additionally shown several literature
descriptions of star loss.
The first one is the analytical description of
\citet[][Eq.~8-85]{BT87} based on the Fokker-Planck approximation 
for the mass loss by evaporation of stars from a cluster (with large
$N$).
\begin{equation}
\label{eq:Mloss}
M(t) = M_{0} \left(1-\frac{7 k_\mathrm{e} t}{2 t_\mathrm{relax}^{0}}\right)^{2/7},
\end{equation}
with $M_{0}$ being the initial cluster mass, $k_\mathrm{e}$ $\approx$
0.003 and $t_\mathrm{relax}^{0}$ the initial relaxation time
(eq.~\ref{eq:trelax}).
With the use of the mean mass, $m_\mathrm{mean}$ of the IMF, which is
for a Kroupa IMF $m_\mathrm{mean}$ $\approx$ 0.36 $M_\odot$, the
number of stars lost over time can be approximated:
\begin{equation}
\label{eq:Nloss}
N(t) = M(t) / m_\mathrm{mean}.
\end{equation}
This equation is indicated with {\it solid black dots} in the two
Figs.~\ref{fig:sres2b} and \ref{fig:sres5b}.

This description is strictly valid only for large $N$ and does not seem
to provide a good characterisation of the $N$-body calculations.
\citet{H74} assumed that in low-$N$ clusters binary interactions
should dominate over the evaporation of stars. He formulated the
following dissolution time scale for low-$N$ clusters:
\begin{equation}
\label{eq:tdiss}
t_\mathrm{diss} = \frac{N^2}{100} t_\mathrm{cross},
\end{equation}
which can also be used to estimate the star loss from a cluster in the
following way:
\begin{equation}
\label{eq:Ntdiss}
N_\mathrm{diss} = N_0 e^{-\frac{\ln(N_0)}{t_\mathrm{diss}}t},
\end{equation}
were $N_0$ is the initial number of stars. This description is included
as {\it open squares} in the Figs.~\ref{fig:sres2b} and
\ref{fig:sres5b}.

An third description of stellar loss can be derived from the
evaporation time scale when assuming the collisionless Boltzmann
equation given in \citet{BT87}. 
\begin{equation}
\label{eq:tevap}
t_\mathrm{evap} = 136 t_\mathrm{relax},
\end{equation}
which can be transformed into the following number loss formula:
\begin{equation}
\label{eq:Ntevap}
N_\mathrm{evap} = N_0 e^{-\frac{\ln(N_0)}{t_\mathrm{evap}}t},
\end{equation}
where $N_0$ is the initial number of stars. In the
Figs.~\ref{fig:sres2b} and \ref{fig:sres5b} this relation is plotted
as {\it open triangles}.

The Figs.~\ref{fig:N1000restr1} to \ref{fig:N1000range} show the results
using the star clusters with initially $N$ = 1000 after 5 Myr while the
Figs.~\ref{fig:N100restr1} to \ref{fig:N100range} show the same for $N$ = 100.

\section{Discussion}
\label{se:dis}

As is visible in Fig.~\ref{fig:N1000restr1}, the number and mass
fraction of lost stars for the N=1000 clusters is only dependent on
the initial radius of the cluster ($\rightarrow$
$t_\mathrm{relax}$), while the binary fraction does not play a major 
role\footnote{It should be noted here that all figures are evaluated after a calculation time of 5 Myr and that the abscissae in most cases is the relaxation time and therefore not a time evolution.}. Likewise in the case of a collapsing cluster (sub-virial, {\it
dashed-dotted line}). A sub-virial cluster just behaves like a
somewhat more concentrated cluster as they collapse and therefore
change their timescales.

In the case of the $N$ = 100 clusters (Fig.~\ref{fig:N100restr1}) there
seems to be a dependence of the number loss on the binary fraction for
objects with large radii ($t_\mathrm{relax}$ $>$ 2 Myr). But
this might still be a low-number stochastic effect as only a couple of
stars leave these clusters ($N_\mathrm{lost}$ $<$ 5). When comparing
the lost mass fraction of the $N$ = 1000 and $N$ = 100 calculations,
the richer cluster seem to loose 2 or 3\% more mass, reflecting
the higher encounter rate in the richer objects. Allowing 
for more massive stars in the low-$N$ clusters ({\it short-dashed
lines} in Figs.~\ref{fig:N100restr1} to \ref{fig:sres5b}) increases
the star loss by similar amounts as a high-binary fraction. This
is probably due to the higher average mass of the systems compared to
single stars. A higher upper mass limit of the IMF likewise increases
the average stellar mass. A larger average mass leads to higher
momenta and therefore to larger velocities after 2- and 3-body
interactions which are then more likely to exceed the escape
velocity of the cluster.

Like the mass loss, the mean escape velocity, $<$$v_\mathrm{esc}$$>$
({\it Panel A} of Fig.~\ref{fig:N1000restr2}), is mainly depended on
$t_\mathrm{relax}$, though the existence of a high-binary fraction
raises $<$$v_\mathrm{esc}$$>$ by about 1 km/s, independent of
$t_\mathrm{relax}$. Primordial binaries introduce 3- and 4-body
interactions early-on in the cluster evolution opposed to the mainly
2-body interactions for single star-only clusters. Such higher order
interactions generally result in larger velocities.

The mean velocity of the ejected stars in the
case of $N$ = 1000 is 4 to 8 km/s (Fig.~\ref{fig:N1000restr2}) and the
escaped stars travel on the average up to 10 to 30 pc
(Fig.~\ref{fig:N1000range}) within 5 Myr depended on the original size
of the cluster\footnote{A velocity of 1 km/s is roughly equal to 1 pc
per Myr.}. But individual stars can be up to 1 kpc away, as the
highest escape velocity encountered for the $N$ = 1000 clusters was
about 260 km/s, though the typical range of escape velocities is
between 2 and 40 km/s. These values are in the same range as the
results of \citet{K98} who studied clusters with $N$ = 400 in a
Galactic tidal field but without a gas background potential. A
comparison with observations seem to contradict the relatively large
number of lost stars found here. \citet{THF09} study the surroundings
of the Orion Nebula cluster (ONC) and find that the majority of stars
with velocities differences to the cluster larger than 10 km/s have
little or no IR-excess and are therefore most likely old and not
associated with the ONC. When looking only at escapees with velocities
larger than 10 km/s after only 1 Myr \citep[the age of the
ONC,][]{HH98}, only between 0 and 14 are to be expected from the
$N$-body calculations, depending on the initial
conditions. Interestingly, \citet{THF09} do find 6 stars with a
velocity offset above 10 km/s which have an IR-excess consistent of
being young stars and therefore could have been ejected from the ONC,
consistent with our predictions. As the \citet{THF09} study can not
differentiate between bound and unbound cluster stars below 10 km/s it
is currently not possible to test the predictions of the $N$-body
calculations further. Also consistent with this picture is the
recently discovered halo of low-mass stars around the small young
cluster $\eta$ Chamaeleontis \citep{MLB10b}. Generally, though, is an
IR-excess probably not a good age indicator for ejected stars as most
of these had a close encounter with another star which lead to the
ejection. During this encounter any disc around that star could be
truncated or largely stripped, therefore reducing or eliminating the
IR-excess. In order to search for a distributed population of ejected
young stars around clusters the position of the stars in the HR
diagram will likely be a more robust age indicator.

For less rich clusters ({\it Panel A} of Fig.~\ref{fig:N100restr2}),
smaller $<$$v_\mathrm{esc}$$>$ are generally achieved. These are also
increased by the presence of binaries by 0.5 to 1 km/s. Stars escaping
from $N$ = 100 clusters on the average travel 3 to 6 pc
(Fig.~\ref{fig:N100range}) within 5 Myr with a typical range of
escape velocities between 1 and 12 km/s. For the low-$N$ clusters
the fastest escaper had about 40 km/s and travelled about 100 pc
in total. Interestingly, the mean of the velocity dispersion of
escape velocities, $\sigma$, is more strongly affected by the presence
of binaries, both in the $N$ = 1000 
({\it Panel B} of Fig.~\ref{fig:N1000restr2}) and $N$ = 100 case ({\it
  Panel B} of Fig.~\ref{fig:N100restr2}). Therefore, the presence of large 
fractions of binaries, while only slightly enlarging the mean escape
velocity, adds more scatter to the velocities of the escaping
stars. This increases the area around a young embedded cluster where
to expect escaping stars significantly by 50 to 200\%.
The mean mass of the escaping stars, however, seems to be rather
independent of binaries in the $N$ = 1000 clusters but is strongly
correlated with $t_\mathrm{relax}$ ({\it Panel A} of Fig.~\ref{fig:sres2b}). 
It rises from about 0.22 $M_\odot$ in the 10 Myr case ($\rightarrow$
$R_\mathrm{ecl}$ = 0.5 pc) to about 0.4 $M_\odot$ for very compact (1 Myr
$\rightarrow$ $R_\mathrm{ecl}$ = 0.1 pc) clusters, while for the
collapsing clusters it stays at about 0.3 $M_\odot$. On the contrary
for the low-$N$ = 100 clusters, $<$$m_\mathrm{lost}$$>$ roughly stays
constant with $t_\mathrm{relax}$ ({\it Panel A} of
Fig.~\ref{fig:sres5b}) but varies strongly with the 
binary  fraction. Here the sub-virial, 50\% binary and higher upper
mass limit case are rather similar and well separated from the
only-single star case and the 100\% binaries.

In appendix~\ref{app:hist} histograms of the mass and
$<$$v_\mathrm{esc}$$>$ of the escaped stars for the different initial
conditions are shown. Mass histograms are only plotted for the cases
without initial binaries for $N$ = 1000 and $N$ = 100 ({\it Panels A}
of Figs.~\ref{fig:sr1} and \ref{fig:sr4}). The mass histograms in the
$N$ = 1000 look very similar to the input IMF and for $N$ = 100
suffer from the low number of escaped stars. Velocity histograms are
only shown for the cases with 100\% initial binaries. All velocity
histograms are also rather similar but the cases with binaries have a
larger high-velocity tail when compared to the single star only run.

\subsection{Comparison with analytical estimates}

The {\it Panels B} of Figs.~\ref{fig:sres2b} and \ref{fig:sres5b}
show the number of stars lost per number of relaxations the
clusters experiences during 5 Myr for $N$ = 1000 and $N$ = 100,
respectively. Besides the results from the different initial
conditions there are also shown the analytic estimates as given by
eqs.~\ref{eq:Nloss} ({\it solid black dots}), \ref{eq:Ntdiss} ({\it
open squares}) and \ref{eq:Ntevap} ({\it open triangles}). 
In both cases the eq.~\ref{eq:Nloss} predicts far too low escape
numbers which is expected as this equation should be applied only to
clusters with large $N$. For the $N$ = 1000 case the resulting star
loss is between the predictions of eqs.~\ref{eq:Ntdiss} and
\ref{eq:Ntevap} but somewhat closer to the binary interaction
dominated model (eq.~\ref{eq:Ntdiss}). This can be seen as 
an indication that star-loss in such clusters is not purely dominated
by binary interactions (eq.~\ref{eq:Ntdiss}) but evaporation
(eq.~\ref{eq:Ntevap}) already plays a role. Though, the pure
evaporation model of eq.~\ref{eq:Ntevap} technically also only applies
to clusters with larger $N$. The sub-virial models are much higher above the
predictions of eq.~\ref{eq:Ntdiss}, probably because the shorter the
time scales of the collapsing clusters lead to more binary
interactions compared to the virial cases.

The lower $N$ clusters with $N$ = 100 behave different in the
sense that the numerical number of escapees is in between the
estimates given by eqs.~\ref{eq:Nloss} and \ref{eq:Ntevap}, but the
binary interaction dominated model (eq.~\ref{eq:Ntdiss})
overestimates star-loss strongly, while the evaporation model
(eq.~\ref{eq:Ntevap}) seems to give a result closer to the numerical
calculations. Here it might be that higher escape velocity due to the
background potential impedes the loss of stars after low-energy
encounters. As the radii for the $N$ = 100 and the $N$ = 1000 clusters
are the same, the more massive clusters have higher densities and
therefore higher encounter rates. For these it is therefore more
likely that they experience high-energy encounters compared to the $N$
= 100 clusters and therefore the $N$ = 1000 clusters might be less
affected by the background potential.

It should be noted here that none of the analytical
estimates take any background potential into account and only
focus on one physical process for star-loss, either energy
equipartition or binary interactions.

\section{Conclusions}

\begin{itemize}
\item Low-mass star clusters loose between 1\% and 20\% of their stars
  (between 1\% and 15\% of their mass) within their first 5 Myr.
\item This is nearly independent whether or not the clusters have
  primordial binaries but collapsing clusters loose more stars.
\item The mean velocities of the escaping stars depend on the richness
  of the cluster and its radius and can be up to 8.5 km/s for $N$ =
  1000 and $R_\mathrm{ecl}$ = 0.1 pc. This is slightly enlarged if
  primordial binaries are present.
\item The velocity dispersion of the escape velocities is about half
  of the mean escape velocity when less than 50\% of the stars are in
  primordial binaries. In the case of high binary fractions it can be
  as large as the mean escape velocity itself.
\item The mean mass of the escapees depends on the richness of the
  cluster, its radius and the presence of binaries and vary up to an
  factor of 4.
\item None of the analytic estimates from the literature describe the
  star loss precisely but they generally give the right order of
  magnitude of lost stars.
\item The currently known number of fast escapees in the ONC is
  consistent with the predictions of the $N$-body calculations.
\end{itemize}

Henceforth, as the majority of the escapees are not fast
run-away stars but evaporate relatively slow, the presence of 10\% to 20\%
of distributed stars in a young star-forming region can be attributed
to dynamical ejections from low-$N$ embedded star clusters. Especially, 
when taking into account that un-resolved binaries and chance
superpositions of stars will tend to underestimate the number of
stars within a cluster \citep{KG08,MA08,MA09,WK07c}. No
such miscounting due to crowding is to be expected outside of the
clusters, leading to a overestimation of the ratio of stars around
clusters to stars inside of them. Current estimates for the
amount of distributed stars in young star-forming regions are
therefore not in contradiction with the assumptions that most
stars are formed in tight embedded clusters, which is consistent
with other $N$-body studies \citep{K98,KB03}.

\section*{Acknowledgements}
We would like to thank the referee Simon Goodwin for valuable
suggestions and comments. This work was financially supported by the
CONSTELLATION European Commission Marie Curie Research Training
Network (MRTN-CT-2006-035890).

\begin{appendix}
\section{Mass and velocity histograms}
\label{app:hist}
\begin{figure}
\begin{center}
\includegraphics[width=8cm]{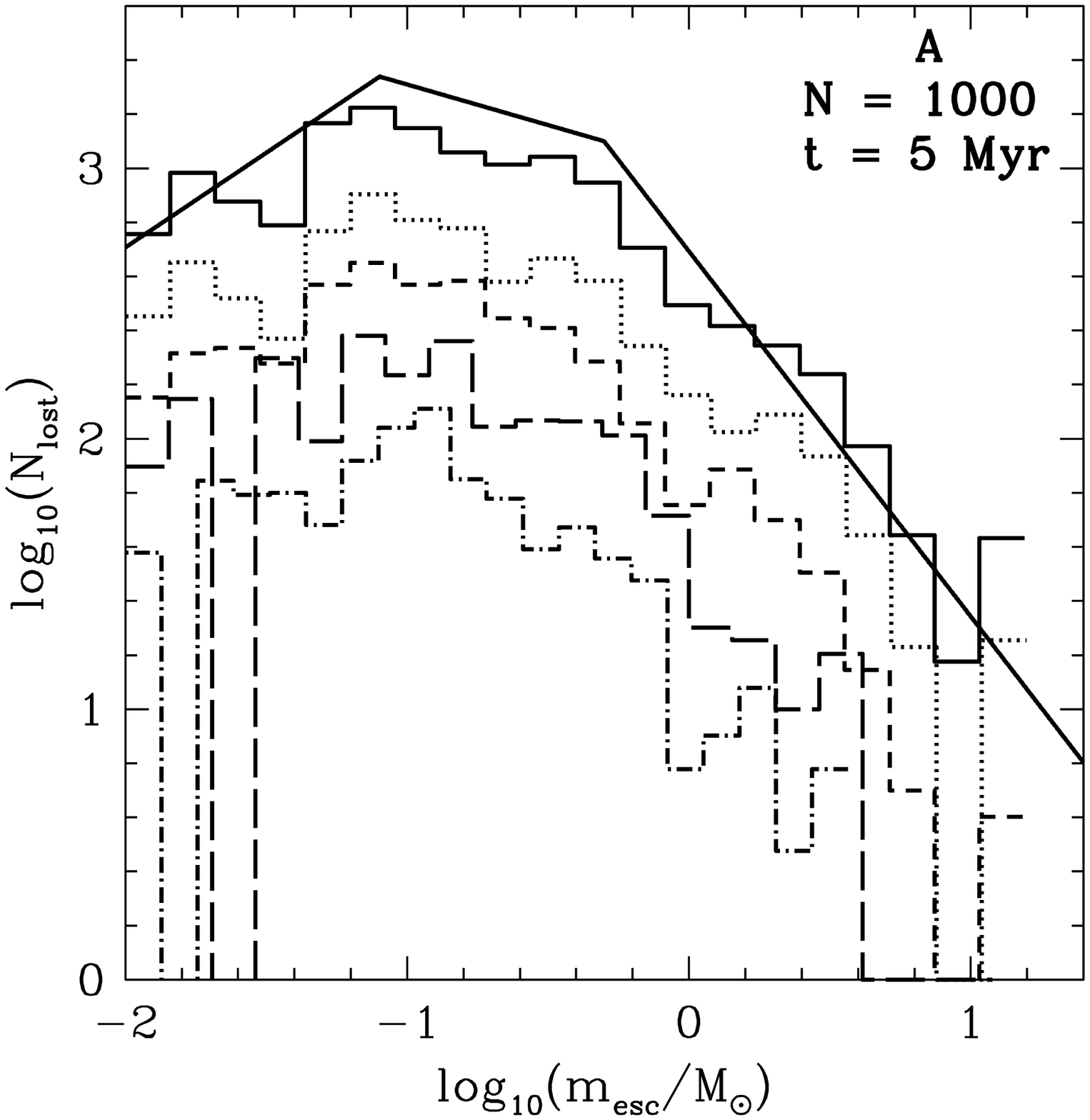}
\includegraphics[width=8cm]{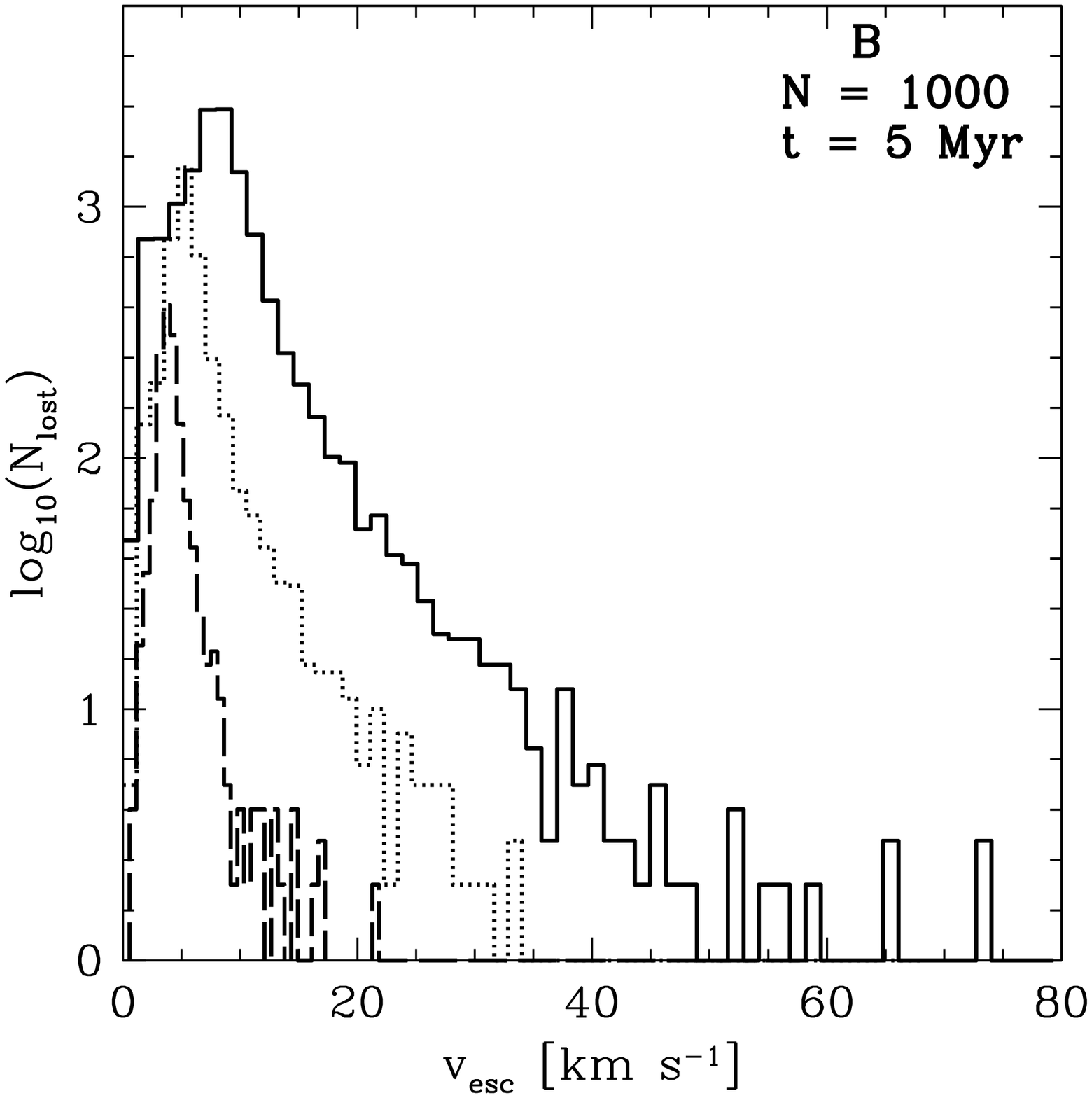}
\vspace*{-2.0cm}
\caption{The mass function ({\it Panel A}) of the lost stars for the
  models started without initial binaries and with a 1000 stars. {\it
  Panel B: } velocity histogram for the case with 100\% initial binaries.
  All models started without initial binaries and 1000
  stars. The {\it solid histogram} is the $R_\mathrm{ecl}$ = 0.1 pc model, 
  the {\it dotted histogram} is the $R_\mathrm{ecl}$ = 0.2 pc model, 
  the {\it short-dashed histogram} is the $R_\mathrm{ecl}$ = 0.3 pc model, 
  the {\it long-dashed histogram} is the $R_\mathrm{ecl}$ = 0.4 pc model and
  the {\it dashed-dotted histogram} is the $R_\mathrm{ecl}$ = 0.5 pc
  model. The {\it solid line} in the mass function histogram is the
  input IMF (not to scale).}
\label{fig:sr1}
\end{center}
\end{figure}

\begin{figure}
\begin{center}
\includegraphics[width=8cm]{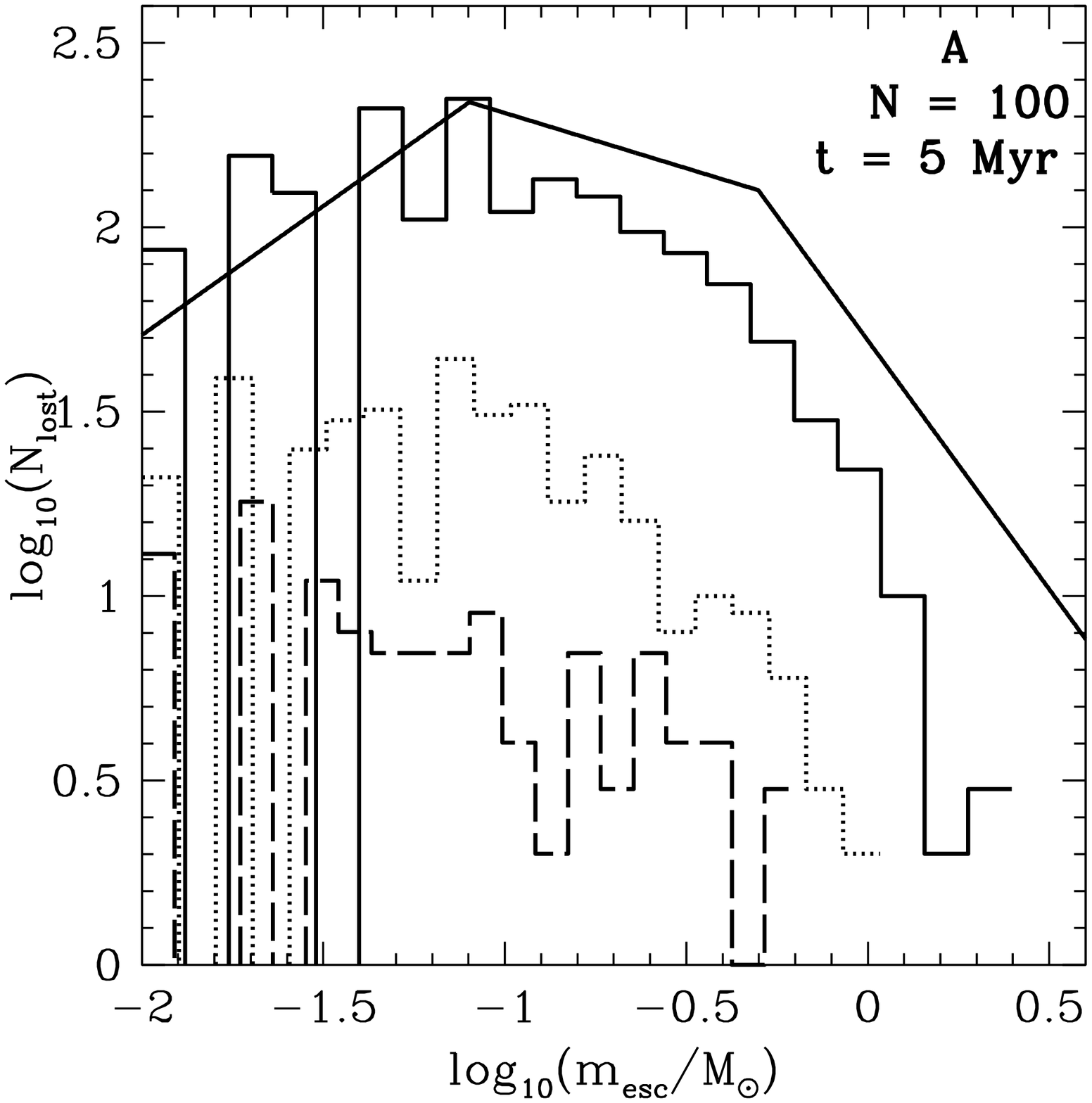}
\includegraphics[width=8cm]{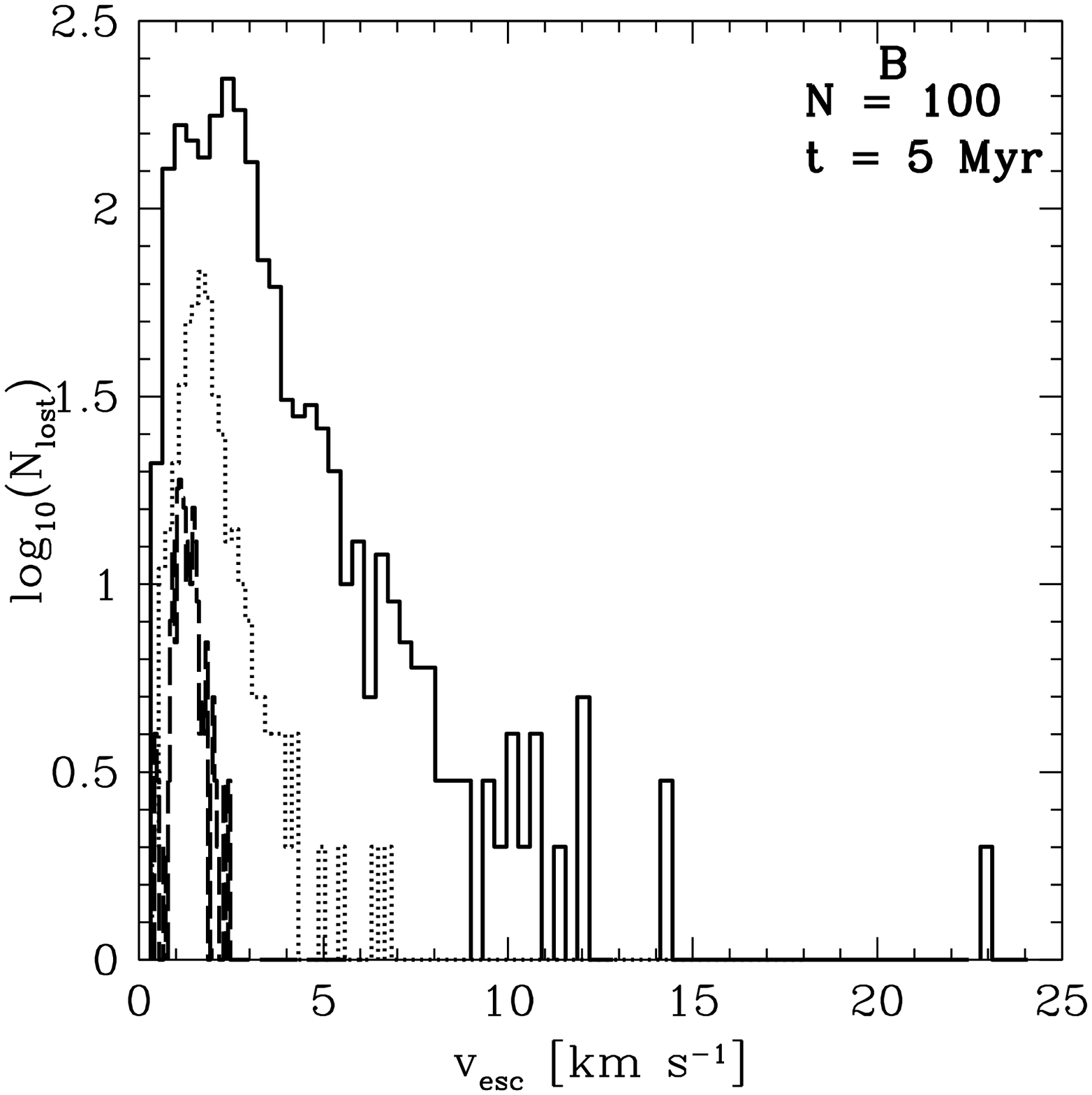}
\vspace*{-2.0cm}
\caption{The mass function ({\it Panel A}) of the lost stars for the
  models started without initial binaries and with a 100 stars. {\it
  Panel B:} velocity histogram for the case with 100\% initial binaries.
  All The {\it solid histogram} is the $R_\mathrm{ecl}$ = 0.1 pc model,
  the {\it dotted histogram} is the $R_\mathrm{ecl}$ = 0.25 pc model and 
  the {\it long-dashed histogram} is the $R_\mathrm{ecl}$ = 0.5 pc
  model. The {\it solid line} in the mass function histogram is the
  input IMF (not to scale).} 
\label{fig:sr4}
\end{center}
\end{figure}

In this appendix are shown the histograms of the mass and velocities
of the escaped stars. Fig.~\ref{fig:sr1} shows a mass histogram and
velocity histogram example for $N$ = 1000 while Fig.~\ref{fig:sr4}
shows the same histograms for $N$ = 100.

\end{appendix}

\bibliography{mybiblio}

\bsp
\label{lastpage}
\end{document}